# Lévy Flight of Photoexcited Minority Carriers in Moderately Doped Semiconductors: Theory and Observation

**Arsen Subashiev and Serge Luryi**
*State University of New York, Stony Brook, NY 11794-2350, U.S.A.*

## 1. Introduction

Diffusive transport of charge carriers in semiconductors has been the subject of detailed studies for at least half a century because of its importance for all electronic and optoelectronic devices.[1-7] These studies have been based on two main approaches. In the first approach, a stationary distribution of minority carriers is created either by optical excitation or an electron beam. One then measures the surface photo-voltage by a capacitance[2] or a Kelvin[3] probe. Alternatively, one registers the photoluminescence spectra[4] or the electron-beam induced currents.[5] These studies give a measure of the diffusion length $l = (D\tau)^{1/2}$, the average distance travelled by minority carriers before they recombine, which is determined from the exponential decay of the distribution with distance from the excitation area. Here $D$ is the diffusion constant and $\tau$ is the carrier lifetime.

The second approach, originating from Haynes and Shockley,[6] uses the time-of-flight measurements,[7] when a short injection pulse (or optical excitation pulse) creates an excess concentration of minority carriers at the edge of a long sample. An external field (the sweep field) across the sample initiates carried drift. Both the drift and the diffusion of carriers are registered by placing two probes along the sample, separated by a certain distance $d$. The drift velocity is measured as $v_D = d/t_0$, where $t_0$ is the time of carrier drift between contacts, measured from pulses in the two probes. The diffusion coefficient is determined through the width at half-maximum of the pulse at the second probe. By this experiment one can validate the Einstein relation between the diffusion coefficient and the carrier mobility.

As demonstrated by these studies, the diffusion equation with a drift term (if an electric field is present) and appropriate boundary conditions can amply describe *with few exceptions* all the experimental results in various semiconductor structures.

The underlying physics in the diffusion process is random motion of particles. If their speed is so large that the effects of the finite time of free flight are not observable, this random motion is equivalent to a random walk with some probability distribution for the size of a single step. Statistical analysis of this type of problems relies on the central limit theorem (CLT). The theorem predicts the distribution function of the final positions of spreading particles after a large number of steps. These positions emerge as a sum of many random steps with an identical single-step probability distribution (SSPD) function, $\mathcal{P}(x)$. According to the CLT, the final distribution should be normal (Gaussian), for any SSPD. The width of the distribution, which gives an average spread of the final positions after $n$ steps, depends on the SSPD. It is given by $\langle x^2 \rangle = \langle x_i^2 \rangle n$, where $\langle x_i^2 \rangle$ is the second moment of the SSPD and $n$ is the number of steps. For a given average temporal interval between the steps $\langle t \rangle$, the spread variance is $\langle x^2 \rangle = 2D\langle t \rangle$, where $D = \langle x_i^2 \rangle / (2\langle t \rangle)$ defines the diffusion coefficient. The linear growth of the spread variance with time is a key feature of the normal diffusive process.

Exceptional cases when the normal diffusion description is not sufficient are of special interest. Besides quantum modifications of the carrier transport, a fairly wide range of classical transport effects that do not follow the normal diffusion pattern have been reported. In this anomalous transport (often referred to as the "anomalous diffusion") the spread variance deviates from the linear growth in time.[8] Clearly, such a deviation must be rooted in a failure of the CLT.

From the definition of the diffusion coefficient $D$ one can identify two possible mechanisms of anomalous kinetics. For some SSPD that slowly decay with the distance, the variance $\langle x_i^2 \rangle$ may become infinitely large. Alternatively, the diffusion is also anomalous when $\langle t \rangle$ is infinitely large. The latter case is typical for the transport kinetics in disordered solids where impurities serve as capture centers for carriers. The random jumps then originate from the capture center activation, and the slowly decaying function is the distribution of waiting times between the jumps. Importantly, in both cases the CLT does not hold: the distribution of a sum of random numbers does not converge to a Gaussian. Still, the resulting distributions for these sums are well understood by the modern probability theory and are known as stable distributions.[9]

Recently, a rather dramatic example of anomalous diffusion was found in the studies of minority-carrier transport in semiconductors with high radiative efficiency,[10-12] when the spread of minority carriers is primarily due to the photon recycling process with the emission-reabsorption events repeated multiple times. The secondary photons, when emitted in the red wing of the emission spectrum, can propagate to large distances, leading to a "heavy-tailed" SSPD, with a power-law decrease with the distance. The resulting anomalous carrier transport features super-diffusive temporal variation and an anomalously wide stationary distribution of the minority carriers. Such transport processes are called the Lévy flight (after Pierre Lévy, the discoverer of the stable distributions).

Here we discuss the results of direct studies[12] of a stationary distribution of the minority carriers, created by an optical excitation in the geometry that combines the advantages of the Haynes-Shockley experiment with those of the photo-luminescence spectroscopy. These results provide an unambiguous demonstration of the Lévy flight transport of holes in moderately doped $n$-InP.

## 2. Anomalous diffusion of the minority carriers

We consider the spatial distribution of holes created by optical excitation of an $n$-doped semiconductor. The energy relaxation time for non-equilibrium minority carriers, created by reabsorption is due to electron-phonon interaction and at room temperature it is in the picosecond range, *i.e.* much shorter than the radiative recombination time. Therefore the temporal evolution of the hole concentration $p(\mathbf{r},t)$ can be studied using the integro-differential transport equation:[10-14]

$$\partial p/\partial t - D\Delta p = -p/\tau + G(\mathbf{r},t) + R(\mathbf{r},t) , \qquad (1)$$

where $G(\mathbf{r},t)$ is the generation function defined as the concentration of holes generated optically per unit time. For a single hole generated at $\mathbf{r} = 0$ at time $t = 0$, this function is $G(\mathbf{r},t) = \delta(\mathbf{r})\delta(t)$. The recombination process is characterized by an average lifetime $\tau$ of holes, which depends on the electron concentration and is in the nanosecond range. It can be either radiative ($\tau_R$) or nonradiative ($\tau_{NR}$), and the rates of these processes are additive, $1/\tau = 1/\tau_R + 1/\tau_{NR}$. The relative probability of radiative recombination is given by the emission quantum efficiency $\eta$:

$$\eta = \frac{\tau_{NR}}{\tau_R + \tau_{NR}} . \qquad (2)$$

The non-radiative lifetime in high-quality crystals reaches several microseconds and $\eta$ can be as high as 99%. Therefore, the emitted photons disappear mainly via interband absorption process, resulting in the generation of a new hole and then a new photon emitted.

The last term in Eq. (1) is the recycling function $R(\mathbf{r},t)$ given by

$$R(\mathbf{r},t) = \frac{\eta}{\tau} \int \mathcal{P}(|\mathbf{r} - \mathbf{r}'|) p(\mathbf{r}',t) \, \mathrm{d}\mathbf{r}' , \qquad (3)$$

which describes the concentration of holes generated per unit time at point $\mathbf{r}$ due to the radiative recombination of holes present in the crystal at the time $t$.

For a finite sample, the solution of Eq. (1) should satisfy the boundary conditions for holes

$$-D \, (\mathrm{d}p/\mathrm{d}n)|_b = (S \times p)|_b \qquad (4)$$

where $S$ is the surface recombination velocity, $b$ stands for the boundary surface, and $\mathrm{d}/\mathrm{d}n$ denotes the derivative along direction normal to the boundary.

3. **Single step probability distribution**

The factor $\mathcal{P}(|\mathbf{r}-\mathbf{r}'|)$ in the integrand of Eq. (3) describes the probability that a hole at $\mathbf{r}'$ generates another hole at $\mathbf{r}$ by the above-described emission-reabsorption process. For the two points separated by the distance $r = |\mathbf{r} - \mathbf{r}'|$ this probability is given by

$$\mathcal{P}(r) = \int \mathcal{N}(E) \frac{\exp[-\alpha(E)r]}{4\pi r^2} \alpha_i(E) \, \mathrm{d}E , \qquad (5)$$

where $\alpha(E)$ is the total absorption coefficient and $\alpha_i(E)$ is the absorption due to interband processes only. The integrand in Eq. (5) is the product of three probabilities: (i) the probability of emission of a photon of energy $E$, described by the normalized emission spectral function $\mathcal{N}(E)$; (ii) the propagation probability of this photon over the distance $r = |\mathbf{r} - \mathbf{r}'|$ (this probability is described by the intensity distribution produced by a unit point source); and (iii) the absorption probability of this photon, described by the factor $\alpha_i(E)$.

It should be noted that for a finite sample one must also consider the boundary conditions for the photons. These would modify the probability in Eq. (5). However, due to the high index contrast between the crystal and the ambient, the angle of total internal reflection is small and the radiation escape cone is narrow. Therefore, we shall further assume that the boundary surfaces are totally reflective for photons. The escape of radiation can be accounted for by an apparent decrease of quantum efficiency.

For moderately doped $n$-type InP the distribution $\mathcal{P}(r)$ can be evaluated using Eq. (5) and the experimentally measured[10-12,15-17] interband absorption coefficient $\alpha_i(E)$ and the residual free-carrier absorption. With the known $\alpha_i(E)$, the spectral density $\mathcal{N}(E)$ of photon emission in a quasi-equilibrium radiative recombination process can be obtained from the thermodynamic relation due to van Roosbroek and Shockley,[18]

$$\mathcal{N}(E) = A\alpha_i(E)E^2 \exp[-E/kT] , \qquad (6)$$

which we shall refer to as the VRS relation. Expression (6) represents the "intrinsic" emission spectrum and it agrees very well with room-temperature luminescence spectra measured from thin epitaxial layers (see, *e.g.* Ref. 19).

Note that even in an infinite space without boundaries the solution of Eq. (1) cannot be factorized into product of functions depending on $x, y,$ and $z$ separately, since the transport processes along perpendicular directions are correlated due to the recycling term. However, it can be studied by starting with a one-dimensional (1D) problem and then finding the point-source distribution through it.[11] To accomplish this one needs the 1D reabsorption probability, which is given by

$$\mathcal{P}(x) = \int \mathcal{P}(r) \mathrm{d}y \mathrm{d}z .$$

Numerical evaluation of $\mathcal{P}(x)$ using experimental absorption spectra of InP has shown[10-11] that for low-doped samples in the entire range of $x$, the dependence is very close to

$$\mathcal{P}(x) = \frac{\gamma x_{\min}^\gamma}{2(x_{\min} + |x|)^{1+\gamma}} , \qquad (7)$$

where $x_{\min} \approx 0.1$ μm is a distance scaling parameter that stems from truncation of the power law at short-distances and normalization of the distribution in Eq. (7). The essential parameter of the distribution is the exponent $\gamma$, called the index of the distribution. For the Urbach-tailed absorption spectrum[17] an approximate analytic expression for the index is:[10-11]

$$\gamma = 1 - \Delta/kT , \qquad (8)$$

where $\Delta$ is the tailing energy. Numerical calculations based on $\alpha(E)$ measured for moderately doped samples show[10-11] that the index varies from $\gamma \approx 0.69$ to $0.64$ in the $N_D = 2$–$6\times10^{17}$ cm$^{-3}$ doping range, so that $\gamma$ decreases slightly as $N_D$ increases.

The empirical values of $\gamma$ are in agreement with Eq. (8), which is valid for a tailing region that is broad enough (compared to $kT$).

The step distribution (7) is *heavy-tailed* for $\gamma < 1$ as in Eq. (8). Its second moment diverges already for $\gamma < 2$, which a hallmark of the anomalous diffusion. Sum of the steps does not follow the CLT and is not Gaussian.

The heavy-tailed distribution arises due to the tailing in the absorption and emission spectra: the photons emitted in the red wing of the emission spectrum can propagate far away from the emission point, making the average square of the reabsorption length infinitely large. This property depends on the shape of the interband absorption spectrum and, in particular, on the width of the tailing region. In direct-gap semiconductors, the interband absorption is often approximated by $\alpha_i(E) \sim (E-E_G)^{1/2}$. This approximation neglects the absorption tails, and hence misses the tailing effect in the reabsorption probability.

Equation (8) correctly predicts the decrease of $\gamma$ with increasing doping level. This effect is due to the smearing of the absorption edge at higher doping, described by increasing tailing energy $\Delta$. Lower values of $\gamma$ are also predicted for lower temperatures.

The residual free-carrier absorption (FCA) leads to the *truncation* of the step distribution (7) at distances $x > 1/\alpha_{FC}$:

$$\mathcal{P}(x) \sim (1/x)^{1+\gamma} \exp[-\alpha_{FC} x] , \qquad (9)$$

where $\alpha_{FC}$ is the FCA coefficient. In *n*-InP one has $\alpha_{FC} \approx 1.3 \times 10^{-18} N_D$, where $\alpha_{FC}$ is in cm$^{-1}$ and $N_D$ is in cm$^{-3}$.

The truncated single-step distribution (9) has a finite second moment and one could expect restoration of the central limit theorem, so that the stationary particle distribution would exponentially decay from the source, as in ordinary diffusion. However, as we shall show both theoretically and experimentally, the truncation does not restore the exponentially decaying stationary distribution.

## 4. Distribution along a band: Boundary effects

Consider a sample in the shape of a long (in $x$ direction) band, whose width ($y$) and thickness ($z$) are both much larger than the ordinary hole diffusion length. Let a stationary photo-excitation be uniform across the short edge face, as shown in Fig. 1. Neglecting the loss of holes or photons at the front and the back broadside surfaces, the hole concentration will remain uniform in the cross-section ($y$, $z$) and the problem becomes one-dimensional. In this geometry, Eq. (1) is of the form

$$-D(\mathrm{d}^2 p/\mathrm{d}x^2) + \frac{p}{\tau} = G(x) + \frac{\eta}{\tau} \int_0^\infty [\mathcal{P}(x-x') + \mathcal{P}(x+x')] \, p(x') \, \mathrm{d}x' , \qquad (10)$$

with the boundary conditions (4) at the $x = 0$ edge (see Fig. 1) and Eqs. (7, 9) for the kernel function. We have assumed total reflection of photons from the edge face surface by including an image source in the integrand. One can see that this effect can be accounted for if one extends both the hole concentration $p(x)$ and the

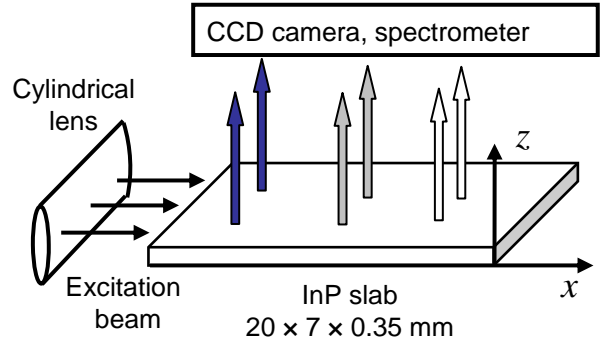

**Figure 1**. Schematic illustration of the experimental geometry.[12] An 808 nm laser excitation beam is focused on a narrow strip in the 7 mm edge of the sample and the photoluminescence is observed from the broadside.

range of integration to the entire axis by setting $p(-x) = p(x)$. This brings Eq. (10) into the form:

$$-D(d^2p/dx^2) + \frac{p}{\tau} = G(x) + \frac{\eta}{\tau}\int_{-\infty}^{\infty} \mathcal{P}(x-x')\, p(x')\, dx' \,. \tag{11}$$

Using the Fourier transformation, one can bring Eq. (11) into algebraic form. We write down the solution for the case of low-doped samples, when the Fourier transform of the kernel function (7) at $k \ll 1/x_{min}$ is given by $\mathcal{P}(k) = 1 - D_\gamma k^\gamma$.

The Fourier transform of the hole concentration $p(k)$ is then given by

$$p(k) = \tau \frac{G(k) - 2Sp(0)}{(1-\eta) + k^2 l^2 + \eta D_\gamma \tau k^\gamma} \,. \tag{12}$$

Here $l = (D\tau)^{1/2}$ is the normal diffusion length, while $D_\gamma$ is the anomalous diffusion coefficient:

$$D_\gamma = \frac{x_{min}}{\Gamma(\gamma)\sin(\pi\gamma/2)\tau}. \tag{13}$$

The spatial hole distribution can be found by the inverse Fourier transform of Eq. (12), which however involves the edge concentration $p(0)$. An equation for $p(0)$ is obtained by applying the result at $x = 0$. In this way, we finally find

$$p(k) = \frac{\tau}{\pi(1 + S\tau J)}\int_0^\infty \frac{G(k)\cos(kx)}{(1-\eta) + k^2 l^2 + \eta D_\gamma \tau k^\gamma}\, dk \,, \tag{14}$$

where

$$J = \frac{2}{\pi}\int_0^\infty \frac{1}{(1-\eta) + k^2 l^2 + \eta D_\gamma \tau k^\gamma}\, dk \,. \tag{15}$$

It follows from Eq. (14) that the surface recombination does not change the hole distribution profile: it manifests itself only through a multiplier factor which can be interpreted as an effective decrease of the excitation intensity due to some losses of holes at the surface. Therefore, we shall omit this factor from this point onwards.

For optical excitation with an absorption coefficient $\alpha_0$, the generation function is of the form $G(x) = I_0\alpha_0\exp[-\alpha_0 x]$ and its Fourier transform is given by

$$G(k) = \frac{I_0\alpha_0^2}{\alpha_0^2 + k^2} \,. \tag{16}$$

From Eq. (16) one can see that when the absorption length $1/\alpha_0$ for the optical excitation is smaller than the diffusion length $l$ of holes, one can approximate the generation function as $G(k) \approx I_0$.

When the quantum radiative efficiency is low, $\eta < 0.5$, the photon recycling is immaterial and Eq. (14) reduces to the well-known solution that describes normal diffusion:

$$p(x) = I_0(\tau/l)\exp[-x/l] \,. \tag{17}$$

Furthermore, if $l \gg (D_\gamma\tau)^{1/\gamma}$, then the spread of holes due to anomalous diffusion is negligible, even though the recycling effect may enhance the effective diffusion length:

$$l_{\text{EFF}} = l(1 - \eta)^{-1/2}. \tag{18}$$

The enhancement is due to the increase of the carrier lifetime by recycling. Of prime interest to us, however, is the opposite case when $(D_\gamma\tau)^{1/\gamma} \gg l$. In this case, the hole distribution profile is dominated by the anomalous diffusion and the normal diffusion term can be omitted for all $x \gg l$, leading to:

$$p(x) \cong \frac{\tau I_0}{\pi(1-\eta)} \int_0^\infty \frac{\cos(kx)}{1 + (x_F k)^\gamma} \, dk \,, \tag{19}$$

where $x_F$ is the excitation "front" distance,

$$x_F = (\Phi \, D_\gamma\tau)^{1/\gamma} \,. \tag{20}$$

Here $\Phi = \eta/(1 - \eta)$ is the recycling factor, *i.e.* the average number $<N>$ of steps in the recycling process; for $\eta \approx 99\%$ one has $\Phi = <N> \approx 100$.

Next let us consider the asymptotics of $p(x)$ at small and large distances. To obtain these asymptotics, one can integrate Eq. (19) by parts. This gives

$$p(x) \cong \frac{\tau I_0 x_F^\gamma \gamma}{\pi(1-\eta)x^{1+\gamma}} \int_0^\infty \frac{\kappa^{\gamma-1}\sin(\kappa)}{(1 + [(x_F/x)\kappa]^\gamma)^2} \, d\kappa \,. \tag{21}$$

It follows that in the limit of large $x \gg x_F$ one has

$$p(x) \cong \frac{\tau I_0 x_F^\gamma \gamma}{\pi(1-\eta)x^{1+\gamma}} \Gamma(1 + \gamma)\sin(\gamma\pi/2) \tag{22}$$

We see that at large distances the distribution is proportional to the probability of a single big jump, multiplied by the recycling factor that reflects the number of attempts available before the jump occurrence.

Conversely, at small distances, $x \ll x_F$, one has from Eq. (21)

$$p(x) \cong \frac{\tau I_0}{\pi(1-\eta)x_F^\gamma x^{1-\gamma}} \Gamma(1-\gamma)\sin(\gamma\pi/2) , \qquad (23)$$

so that the decay of concentration still follows a power law, but with a smaller exponent. The rapid decay (21) of the concentration begins at $x \approx x_F$, which can hence be regarded as the distance to the excitation front characterizing the particle spread. In contrast to the normal diffusion, for which the particle spread is proportional to $<N>^{1/2}$, for the Lévy flight process the excitation front distance (20) grows as $<N>^{1/\gamma}$, which for $\gamma < 1$ and large $<N>$ can be many orders of magnitude larger.

The experimental results,[12] described in the next Section, show that the observed effect is huge: the spread of holes is increased from a few microns (characteristic of ordinary diffusion) to several millimeters. The effect can be further drastically increased by lowering the temperature.

## 5. Experimental luminescence spectra and observation of the Lévy flight distribution of the minority carriers

The geometry of the experiment[12] was shown in Fig. 1. The photoluminescence was optically excited by an above-bandgap laser (808 nm) at the edge face of an InP slab and both the luminescence radiation spectra and luminescence intensity were observed from the broadside and measured as a function of the distance $x$ from the edge. The intensity distribution was obtained by scanning of the CCD image.

The observed intensity was proportional to the excitation intensity. Apart from the normalization, the luminescence spectra were identical for all distances $x > d$. Moreover, their shape agreed with the calculated *filtered* spectra,

$$S(E) = F_1(E)T(E)\mathcal{N}(E) , \qquad (24)$$

where $F_1(E)$ is a one-pass filtering function that describes wavelength-dependent attenuation of the luminescence by sample absorption on the way out,

$$F_1(E) = \int_0^d p(z)\exp[-\alpha(E)z]\,dz, \qquad (25)$$

and $T(E)$ is a factor that accounts for multiple surface reflections. The function in Eq. (24) can be easily calculated assuming a uniform hole distribution across the sample,

$$p(x,z)\big|_{x>d} = p(x) . \qquad (26)$$

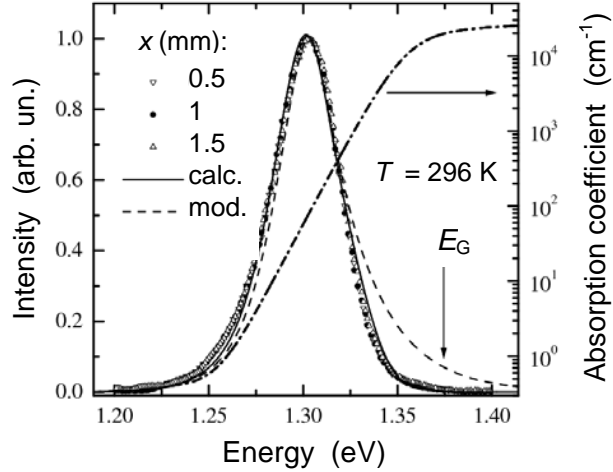

**Figure 2**. Spectra of luminescence observed from the broadside at varying distances from the edge. Solid line shows the spectrum evaluated according to Eqs. (23) and (24) with an accurate account for surface recombination in the calculated $p(z)$ profile. Dashed line corresponds to a simplified model that neglects surface recombination. Also shown is the absorption spectrum, exhibiting a clear Urbach-tail behavior below the bandgap energy.

The experimental spectra are shown in Fig. 2 for several distances $x$ from the edge. We also show the spectrum calculated using Eq. (23) as well as using a simplified model that ignores surface recombination (solid and dashed lines, respectively). The simplified model gives higher intensity in the blue wing and lower intensity in the red wing of the spectrum. More accurate calculations, taking into account the decrease of the hole concentration in a narrow region near the surface, give excellent agreement with the experiment for the entire spectrum.

Near the emission maximum, the variation of the absorption coefficient is well reproduced by the Urbach exponent, and multiple reflections are negligible, so that one can take $T=1$ and set $p(z)$ to a constant value. The position of the observed maximum can be then evaluated by setting $dS(E)/dE = 0$, which gives $E_{MAX} = 1.303$ eV, in a very good agreement with experiment.

The excellent agreement of the calculated spectrum with the experiment and the independence of the spectral shape from $x$ is strong evidence for the proportionality of the luminescence intensity $I(x)$ to the local hole concentration $p(x)$. Note that the luminescence escapes into a narrow cone normal to the surface, therefore the observed $I(x)$ corresponds to local concentration inside the wafer.

The observed intensity distribution $I(x)$ is shown in Fig. 3 (in log-log scale) for two samples with different doping. One can see a huge enhancement of the spread of the carrier concentration – extending to several millimeters – compared with a micron-range spread typical for the normal hole diffusion.[4] For the low-doped sample, a power-law decay is clearly seen at distances $x > 300$ μm. A power-law fit for the index γ gives ~0.7, which agrees well with the theoretical estimate (8). The power law is dramatically distinguishable from an exponential decay (17) expected for normal diffusion – irrespective of any enhancement of the diffusion length (a diffusive curve for $l = 210$ μm is shown for comparison by a dashed line).

For the heavily doped sample, the power law decay is truncated at large distances, $x > 1000$ μm, while the spread (the distance to the excitation front) remains anomalously large. This truncation of the distribution is due to the residual free-carrier absorption, which truncates the SSPD, see Eq. (9). Solid lines show the results of calculations, using Eq. (18) for the low-doped sample and a more accurate solution of Eq. (11) that accounts for the residual absorption in the heavily-doped sample. Both are in a very good agreement with experiment. Note that in spite of the truncation (that provides a finite diffusivity), the stationary distribution reproduces at large distances the single-step probability distribution (9). This is typical for anomalous diffusion.

## 6. A single big jump approximation for the truncated Lévy flight

The residual free carrier absorption that truncates the SSPD also makes its variance finite. Thus, one can use Eq. (9) to calculate a photon-assisted contribution to the diffusion coefficient, $D_{PH} = \langle x_i^2 \rangle / 2\tau$. However, it is clear that when the truncation modifies SSPD only at very large distances (as is the case for truncation by FCA), then its effect on the experimental carrier distribution can hardly be dominant. More detailed analysis of the truncation shows that, for a slowly decaying step distribution, convergence of the sum of $N$ steps to a Gaussian distribution can be very slow. It requires many more steps than the average, $N \gg \langle N \rangle$ and the range of the distances from the source, in which the distribution is close to Gaussian, depends on $N$. Therefore at large distances from the source the distribution can be far from Gaussian even for the truncated SSPD. As an illustration, consider the probability of reaching a large distance $x$ in two steps:[20]

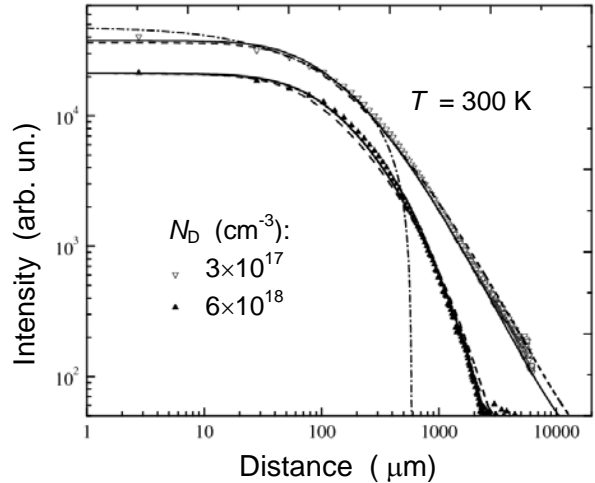

**Figure 3**. Luminescence intensity and the hole concentration $p(x)$ in $n$-InP samples of different $N_D$ vs. distance from the sample edge (presented on a log-log scale). Solid lines show the theoretical curves obtained by solving the integro-differential transport Eq. (11). For the lightly-doped sample, $p(x)$ follows an asymptotic power law with $\gamma = 0.7$. For the heavily-doped sample, the distribution is truncated at large distances ($x > 1$ mm) by free-carrier absorption. Dashed lines show "the single big jump" approximation, Eq. (32); dashed-dotted line shows the best fit by an exponential dependence (17) with an enhanced diffusion length $l = 210$ μm.

$$\mathcal{P}_2(x) = \int \mathcal{P}(|x-x'|)\mathcal{P}(|x'|)dx'. \qquad (27)$$

The integrand of Eq. (27) has two factors, sharply peaked at $x' = 0$ and $x' = x$, respectively. At the peak of one factor, the other factor is declining with a gentle slope, which is true for both the power-law (7) and the exponentially-truncated (9) SSPD. In the limit $x \to \infty$, we can take the slowly declining factor outside the integration with the value it has at the peak of the other factor. This gives

$$\mathcal{P}_2(x) \approx 2\mathcal{P}(x)\int \mathcal{P}(|x'|)dx' . \qquad (28)$$

Equation (28) implies that the most probable way to reach a large distance in two jumps is to do it with one big jump, followed or preceded by a small jump. This is to be contrasted with a rapidly declining SSPD, where the maximum probability would correspond to both steps being of similar size, *i.e.* close to $x/2$. Furthermore, there is a rather broad class of distributions,[20] including the truncated power-law distribution (9), for which the probability of reaching a *large enough* distance $x$ in $n$ steps is close to

$$\mathcal{P}_n(x) \approx n\mathcal{P}(x)[\int \mathcal{P}(|x'|)dx']^{n-1} . \qquad (29)$$

Equation (29) implies that at very large distances $x$, a single big jump dominates the probability to reach $x$ even in $n$ steps. The mathematical theorem corresponding to Eq. (29) can be proven by induction, using (28).

The theorem (29) suggests a way of solving Eq. (11) for a heavy-tailed SSDP – by sequential iterations in the asymptotic region at large distances from the source – making the "longest step approximation" (29) in all terms of the iterative series. This corresponds to choosing one of the steps to equal the total distance, while keeping all the other steps small.

In Eqs. (28) and (29) the integration limits are not important and can be extended to infinity since the integrals converge at distances much shorter than $x$. One can, however, choose the limits in such a way that a good approximation is obtained at small distances too. To do this, we replace the integral in (29) by:

$$J = \int \mathcal{P}(|x'|)dx' = 2\int_0^{cx} \mathcal{P}(x')dx' , \qquad (30)$$

and determine the constant $c$ by sewing together the solutions at small and large $x$.

Using Eqs. (29) and (30), we can sum the iterative series obtained from (11) as a geometric progression, to find

$$p(x) = \eta \mathcal{P}/(1 - \eta J)^2 , \tag{31}$$

or

$$p(x) = \frac{G\tau}{(1 - \eta)} \frac{\Phi \mathcal{P}(x)}{[1 + \Phi \mathcal{P}_{\mathrm{ESC}}(x)]^2} , \tag{32}$$

where

$$\mathcal{P}_{\mathrm{ESC}}(x) = 2\int_{cx}^{\infty} \mathcal{P}(x')\mathrm{d}x' \tag{33}$$

is the probability of escape beyond $cx$ in one step. To find the sewing constant $c$, we specialize to the power-law step distribution (7), where we have

$$\mathcal{P}_{\mathrm{ESC}}(x) = \left(\frac{x_{\min}}{cx}\right)^{\gamma} . \tag{34}$$

It follows from Eqs. (32-34) that the distribution $p(x)$ asymptotically (at large $x$) reproduces the one-step probability (7) or (9), enhanced by the recycling factor $\Phi$. It is modified at small $x$, when the denominator in (32) becomes large. At these small distances the residual absorption in Eq. (33) can be neglected, *i.e.* the approximation (29) is justified. Comparing (32) with the solution (23) valid at small $x$, we find

$$c = \left(\frac{\Gamma(1 + \gamma)}{\Gamma(1 - \gamma)}\right)^{1/2\gamma} \approx \left(\frac{1}{1 - \gamma}\right)^{1/2\gamma} . \tag{35}$$

For $\gamma = 0.7$, Eq. (35) yields $c = 2.4$.

The escape probability (33) allows to further interpret the excitation "front" distance $x_F$ defined by Eq. (20):

$$\Phi \mathcal{P}_{\mathrm{ESC}}(x_F) = 1 . \tag{36}$$

Since the left-hand side of (36) corresponds to the escape probability in $\Phi$ attempts, the distance defined by Eq. (36) is the distance beyond which the holes appear predominantly in one step (the recycling factor $\Phi$ is the average number of attempts in the recycling process). The position of the front is clearly seen in Fig. 3 as the point of maximum curvature on the log-log plots of the hole distribution $p(x)$.

Equation (32) gives a good approximation to the exact solution in the entire range of $x$ – with or without truncation of the Lévy flight (see Fig. 3). The observed values of $x_F \approx 200$ μm are indicative of the large recycling factor $\Phi$ and therefore of the very high radiative efficiency $\eta$ in our InP samples.

## 7. Conclusions

We have studied the anomalous hole distribution arising in photoluminescence experiments in moderately doped *n*-InP samples due to the photon recycling effect. The anomalous distribution manifests itself by a heavy-tailed spread of the holes over large distances from the excitation. It is accurately described by the kinetics of the Lévy flight. The luminescence spectra remain identical at all distances and in good agreement with the spectra of equilibrium hole emission.

The anomalous distributions remain qualitatively similar for heavier doped samples, when the power-law reabsorption probability is exponentially truncated by free-carrier absorption. The experimental results are well described by the theory and can be accurately interpreted using the single big jump approximation.

We believe that the discovered anomalous transport of minority carriers in semiconductors will have a large impact on the design of a number of optoelectronic devices such as multicolored LEDs, opto-thyristors, photovoltaic devices with high efficiency, as well as on the development of a semiconductor scintillator for radiation detection. In all these devices the recycling process is known to be of importance and even was accounted for both in estimations and in the design. As we have shown, it is important to accurately account for an anomalous character of the minority carrier transport. Large spread of the minority carriers opens a possibility of a design for the devices with optimized photonically enhanced carrier transport.

From a broader perspective, there are only a few systems in the nature where Lévy flights can be directly observed in experimental settings with controlled parameters. In the semiconductor system all parameters of the process are well controlled and may be varied (by temperature, doping and the choice of the semiconductor material or structure). This makes it a unique object for the studies. Moreover, there is a rare possibility to experimentally control and accurately identify the truncation mechanism.

**Acknowledgments**

This work was supported by the Domestic Nuclear Detection Office, by the Defense Threat Reduction Agency (basic research program), and by the Center for Advanced Sensor Technology at Stony Brook.

**References**

1. M. Lundstrom, *Fundamentals of Carrier Transport*, 2nd ed., Cambridge, UK: Cambridge University Press, 2000.
2. D. K. Schroder, "Surface voltage and surface photovoltage: History, theory and applications," *Meas. Sci. Technol.* **12**, R16 (2001).
3. R. Shikler, N. Fried, T. Meoded, and Y. Rosenwaks, "Measuring minority-carrier diffusion length using a Kelvin probe force microscope," *Phys. Rev. B* **61**, 2550 (2000).
4. C. J. Hwang, "Optical properties of n-type GaAs. I. Determination of hole diffusion length from optical absorption and photoluminescence measurements," *J. Appl. Phys.* **40**, 3731 (1969).

5. C. J. Wu and D. B. Wittry, "Investigation of minority-carrier diffusion lengths by electron bombardment of Schottky barriers," *J. Appl. Phys.* **49**, 2827 (1978).
6. J. R. Haynes and W. Shockley, "The mobility and life of injected holes and electrons in germanium," *Phys. Rev.* **81**, 835 (1951); *ibid.* **75**, 691 (1949).
7. A. Sconza, G. Galet, and G. Torzo, "An improved version of the Haynes–Shockley experiment with electrical or optical injection of the excess carriers," *Am. J. Phys.* **68**, 80 (2000).
8. R. Metzer and J. Klafter, "The restaurant at the end of the random walk: Recent developments in the description of anomalous transport by fractional dynamics," *J. Phys. A: Math. Gen.* **37**, R161 (2004).
9. V. M. Zolotarev, *One-Dimensional Stable Distributions*, Providence, RI: American Mathematical Society, 1986.
10. O. Semyonov, A. V. Subashiev, Z. Chen, and S. Luryi, "Photon assisted Lévy flights of minority carriers in *n*-InP," *J. Luminescence* **132**, 1935 (2012).
11. S. Luryi and A. Subashiev, "Lévy flight of holes in InP semiconductor scintillator," *Intern. J. High Speed Electronics Syst.* **21**, 3 (2012); also online *http://arxiv.org/abs/1202.5576v1*
12. S. Luryi, O. Semyonov, A. Subashiev, and Z. Chen, "Direct observation of Lévy flight of holes in bulk *n*-InP," *Phys. Rev. B* **86**, 201201 (2012); also online: *http://arxiv.org/abs/1205.4975*
13. T. Holstein, "Imprisonment of resonance radiation in gases," *Phys. Rev.* **72**, 1212 (1947); L. M. Biberman, "On the diffusion theory of resonance radiation," [in Russian] *Zh. Eksper. Teor. Fiz.* **17**, 416 (1947).
14. V. V. Ivanov, *Transfer of Radiation in Spectral Lines*, Nat. Bureau Standards pub. no. 385, 1973; G. Rybicki and A. Lightman, *Radiation Processes in Astrophysics*, Wiley: New York, 1979.
15. S. Luryi and A. Subashiev, "Semiconductor scintillator for 3-dimensional array of radiation detectors," chapter in: S. Luryi, J. M. Xu, and A. Zaslavsky, eds., *Future Trends in Microelectronics: From Nanophotonics to Sensors to Energy*, New York: Wiley, 2010, pp. 331–346.
16. O. Semyonov, A. V. Subashiev, Z. Chen, and S. Luryi, "Radiation efficiency of heavily doped bulk *n*-InP semiconductor," *J. Appl. Phys.* **108**, 013101 (2010).
17. A. Subashiev, O. Semyonov, Z. Chen, and S. Luryi, "Urbach tail studies by luminescence filtering in moderately doped bulk InP," *Appl. Phys. Lett.* **97**, 181914 (2010).
18. W. van Roosbroek and W. Shockley, "Photon-radiative recombination of electrons and holes in germanium," *Phys. Rev.* **94**, 1558 (1954).
19. R. M. Sieg and S. A. Ringel, "Reabsorption, band-gap narrowing, and the reconciliation of photoluminescence spectra with electrical measurements for epitaxial n-InP," *J. Appl. Phys.* **80**, 448 (1996).
20. S. Foss, D. Korshunov, and S. Zachary, *An Introduction to Heavy-Tailed and Subexponential Distributions*, New York: Springer, 2011.